\documentclass[12pt]{article}
\pdfoutput=1

\usepackage{amsmath,amsthm,amssymb,euscript,epsf,epsfig}
\usepackage{array,latexsym,amsfonts,bm,enumerate,url}

\usepackage{graphicx}
\usepackage[usenames]{color}
\usepackage[english]{babel}

\usepackage{fancybox}

\usepackage{mciteplus}

\usepackage{parskip}     
\usepackage{titlesec}
\titlespacing{\section}{0pt}{3.5ex}{1ex}
\titlespacing{\subsection}{0pt}{2.5ex}{1ex}
\titlespacing{\subsubsection}{0pt}{2ex}{0ex}

\usepackage[
      colorlinks=true,
      linkcolor=blue,
      urlcolor=blue,
      filecolor=blue,
      citecolor=blue,
      pdfstartview=FitH,
      pdftitle={},
      pdfauthor={},
      pdfsubject={},
      pdfkeywords={},
      pdfpagemode=UseNone,
      bookmarksopen=true
      ]{hyperref}
\usepackage[all]{hypcap}     

\begin{document}

\numberwithin{equation}{section}


\mathchardef\mhyphen="2D


\newcommand{\be}{\begin{equation}}
\newcommand{\ee}{\end{equation}}
\newcommand{\bea}{\begin{eqnarray}\displaystyle}
\newcommand{\eea}{\end{eqnarray}}
\newcommand{\nnm}{\nonumber}
\newcommand{\nn}{\nonumber}

\def\eq#1{(\ref{#1})}
\newcommand{\secn}[1]{Section~\ref{#1}}

\newcommand{\tbl}[1]{Table~\ref{#1}}
\newcommand{\fig}{Fig.~\ref}

\def\beq{\begin{equation}}
\def\eeq{\end{equation}}
\def\beqa{\begin{eqnarray}}
\def\eeqa{\end{eqnarray}}
\def\bet{\begin{tabular}}
\def\eet{\end{tabular}}
\def\bs{\begin{split}}
\def\es{\end{split}}


\def\a{\alpha}  
\def\b{\beta}  
\def\c{\chi}
\def\g{\gamma}
\def\G{\Gamma}
\def\e{\epsilon}
\def\vep{\varepsilon}
\def\tvep{\widetilde{\varepsilon}}
\def\f{\phi}
\def\F{\Phi}
\def\fb{{\ov \phi}}
\def\vf{\varphi}
\def\m{\mu}
\def\mub{\ov \mu}
\def\n{\nu}
\def\nub{\ov \nu}
\def\o{\omega}
\def\O{\Omega}
\def\r{\rho}
\def\k{\kappa}
\def\kab{\ov \kappa}
\def\s{\sigma}
\def\t{\tau}
\def\th{\theta}
\def\sb{\ov\sigma}
\def\S{\Sigma}
\def\l{\lambda}
\def\L{\Lambda}
\def\p{\psi}


\def\cA{{\cal A}} \def\cB{{\cal B}} \def\cC{{\cal C}}
\def\cD{{\cal D}} \def\cE{{\cal E}} \def\cF{{\cal F}}
\def\cG{{\cal G}} \def\cH{{\cal H}} \def\cI{{\cal I}}
\def\cJ{{\cal J}} \def\cK{{\cal K}} \def\cL{{\cal L}}
\def\cM{{\cal M}} \def\cN{{\cal N}} \def\cO{{\cal O}}
\def\cP{{\cal P}} \def\cQ{{\cal Q}} \def\cR{{\cal R}}
\def\cS{{\cal S}} \def\cT{{\cal T}} \def\cU{{\cal U}}
\def\cV{{\cal V}} \def\cW{{\cal W}} \def\cX{{\cal X}}
\def\cY{{\cal Y}} \def\cZ{{\cal Z}}

\def\mC{\mathbb{C}} 
\def\mP{\mathbb{P}}  
\def\mR{\mathbb{R}} 
\def\mZ{\mathbb{Z}} 
\def\mT{\mathbb{T}} 
\def\mN{\mathbb{N}}
\def\mH{\mathbb{H}}
\def\mX{\mathbb{X}}

\def\C{\mathbb{C}}
\def\CP{\mathbb{CP}}
\def\R{\mathbb{R}}
\def\RP{\mathbb{RP}}
\def\Z{\mathbb{Z}}
\def\N{\mathbb{N}}
\def\H{\mathbb{H}}

\newcommand{\rmd}{\mathrm{d}}
\newcommand{\rmx}{\mathrm{x}}

\def\tA{ {\widetilde A} } 

\def\one{{\hbox{\kern+.5mm 1\kern-.8mm l}}}
\def\zero{{\hbox{0\kern-1.5mm 0}}}


\newcommand{\red}[1]{{\color{red} #1}}
\newcommand{\blue}[1]{{\color{blue} #1}}
\newcommand{\green}[1]{{\color{green} #1}}

\definecolor{orange}{rgb}{1,0.5,0}
\newcommand{\orange}[1]{{\color{orange} #1}}


\newcommand{\bra}[1]{{\langle {#1} |\,}}
\newcommand{\ket}[1]{{\,| {#1} \rangle}}
\newcommand{\braket}[2]{\ensuremath{\langle #1 | #2 \rangle}}
\newcommand{\Braket}[2]{\ensuremath{\langle\, #1 \,|\, #2 \,\rangle}}
\newcommand{\norm}[1]{\ensuremath{\left\| #1 \right\|}}
\def\corr#1{\left\langle \, #1 \, \right\rangle}
\def\vac{|0\rangle}


\def\d{ \partial } 
\def\zb{{\bar z}}

\newcommand{\sq}{\square}
\newcommand{\IP}[2]{\ensuremath{\langle #1 , #2 \rangle}}    

\newcommand{\floor}[1]{\left\lfloor #1 \right\rfloor}
\newcommand{\ceil}[1]{\left\lceil #1 \right\rceil}

\newcommand{\dbyd}[1]{\ensuremath{ \frac{\d}{\d {#1}}}}
\newcommand{\ddbyd}[1]{\ensuremath{ \frac{\d^2}{\d {#1}^2}}}

\newcommand{\Zd}{\ensuremath{ Z^{\dagger}}}
\newcommand{\Xd}{\ensuremath{ X^{\dagger}}}
\newcommand{\Ad}{\ensuremath{ A^{\dagger}}}
\newcommand{\Bd}{\ensuremath{ B^{\dagger}}}
\newcommand{\Ud}{\ensuremath{ U^{\dagger}}}
\newcommand{\Td}{\ensuremath{ T^{\dagger}}}

\newcommand{\T}[3]{\ensuremath{ #1{}^{#2}_{\phantom{#2} \! #3}}}		

\newcommand{\tr}{\operatorname{tr}}
\newcommand{\Str}{\operatorname{Str}}
\newcommand{\sech}{\operatorname{sech}}
\newcommand{\Spin}{\operatorname{Spin}}
\def\Tr{{\rm Tr\, }} 
\newcommand{\Sym}{\operatorname{Sym}}
\newcommand{\Com}{\operatorname{Com}}
\def\adj{\textrm{adj}}
\def\id{\textrm{id}}
\def\Id{\textrm{Id}}
\def\ind{\textrm{ind}}
\def\Dim{\textrm{Dim}}
\def\End{\textrm{End}}
\def\Res{\textrm{Res}}
\def\Ind{\textrm{Ind}}
\def\ker{\textrm{ker}}
\def\im{\textrm{im}}
\def\sgn{\textrm{sgn}}
\def\ch{\textrm{ch}}
\def\STr{\textrm{STr}}
\def\Sym{\textrm{Sym}}

\def\ha{\frac{1}{2}}
\def\tha{\tfrac{1}{2}}
\def\wt{\widetilde}
\def\ra{\rangle}
\def\la{\langle}

\def\pb{\ov\psi}
\def\pt{\widetilde{\psi}}
\def\at{\widetilde{\a}}
\def\cb{\ov\chi}
\def\d{\partial}
\def\db{\bar\partial}
\def\delb{\bar\partial}
\def\dbar{\ov\partial}
\def\dag{\dagger}
\def\dalpha{{\dot\alpha}}
\def\dbeta{{\dot\beta}}
\def\dgamma{{\dot\gamma}}
\def\ddelta{{\dot\delta}}
\def\ad{{\dot\alpha}}
\def\bd{{\dot\beta}}
\def\dg{{\dot\gamma}}
\def\dd{{\dot\delta}}
\def\th{\theta}
\def\Th{\Theta}
\def\eb{{\ov \epsilon}}
\def\gb{{\ov \gamma}}
\def\wb{{\ov w}}
\def\Wb{{\ov W}}
\def\ib{{\ov i}}
\def\jb{{\ov j}}
\def\kb{{\ov k}}
\def\mb{{\ov m}}
\def\nb{{\ov n}}
\def\qb{{\ov q}}
\def\Qb{{\ov Q}}
\def\xh{\hat{x}}
\def\D{\Delta}
\def\DD{\Delta^\dag}
\def\Db{\ov D}
\def\M{{\cal M}}
\def\rd{\sqrt{2}}
\def\ov{\overline}
\def\Slash{\, / \! \! \! \!}
\def\dslash{\partial\!\!\!/} 
\def\Dslash{D\!\!\!\!/\,\,}
\def\fslash#1{\slash\!\!\!#1}
\def\Fslash#1{\slash\!\!\!\!#1}

\def\del{\partial}
\def\delb{\bar\partial}
\newcommand{\ex}[1]{{\rm e}^{#1}} 
\def\ii{{i}}

\renewcommand{\theequation}{\thesection.\arabic{equation}}
\newcommand{\vs}[1]{\vspace{#1 mm}}

\newcommand{\ve}{{\vec{\e}}}
\newcommand{\shalf}{\frac{1}{2}}

\newcommand{\lb}{\rangle}
\newcommand{\al}{\ensuremath{\alpha'}}
\newcommand{\ap}{\ensuremath{\alpha'}}

\newcommand{\bean}{\begin{eqnarray*}}
\newcommand{\eean}{\end{eqnarray*}}
\newcommand{\ft}[2]{{\textstyle {\frac{#1}{#2}} }}

\newcommand{\hsp}{\hspace{0.5cm}}
\def\half{{\textstyle{1\over2}}}
\let\ci=\cite \let\re=\ref
\let\se=\section \let\sse=\subsection \let\ssse=\subsubsection

\newcommand{\dpb}{D$p$-brane}
\newcommand{\dpbs}{D$p$-branes}

\def\gh{{\rm gh}}
\def\sgh{{\rm sgh}}
\def\NS{{\rm NS}}
\def\R{{\rm R}}
\def\Qp{Q_{\rm P}}
\def\QP{Q_{\rm P}}

\newcommand\dott[2]{#1 \! \cdot \! #2}

\def\eo{\overline{e}}



\def\p{\partial}
\def\h{{1\over 2}}

\def\d{\partial}
\def\la{\lambda}
\def\eps{\epsilon}
\def\bb{\bigskip}
\def\mm{\medskip}
\def\tg{\widetilde\gamma}
\newcommand{\dm}{\begin{displaymath}}
\newcommand{\edm}{\end{displaymath}}
\renewcommand{\b}{\widetilde{B}}
\newcommand{\gm}{\Gamma}
\newcommand{\ac}[2]{\ensuremath{\{ #1, #2 \}}}
\renewcommand{\ell}{l}
\newcommand{\z}{\ell}
\newcommand{\newsection}[1]{\section{#1} \setcounter{equation}{0}}
\def\bb{$\bullet$}
\def\Qbar{{\bar Q}_1}
\def\QPbar{{\bar Q}_p}

\def\q{\quad}

\def\bn{B_\circ}

\let\a=\alpha \let\b=\beta \let\g=\gamma \let\d=\delta \let\e=\epsilon
\let\c=\chi \let\th=\theta  \let\k=\kappa
\let\l=\lambda \let\m=\mu \let\n=\nu \let\x=\xi \let\r=\rho
\let\s=\sigma \let\t=\tau
\let\vp=\varphi \let\vep=\varepsilon
\let\w=\omega      \let\G=\Gamma \let\D=\Delta \let\Th=\Theta
                     \let\P=\Pi \let\S=\Sigma

\def\h{{1\over 2}}
\def\t{\widetilde}
\def\r{\rightarrow}
\def\nn{\nonumber\\}
\let\bm=\bibitem
\def\Kt{{\widetilde K}}
\def\b{\bigskip}

\let\p=\partial


%
%
\begin{flushright}
\end{flushright}
\vspace{2cm}
\begin{center}
{\LARGE  Momentum-carrying waves }\\
\vspace{.5cm}
{\LARGE  on D1-D5 microstate geometries}\\
\vspace{2cm}

{\bf Samir D. Mathur ~and~ David Turton} \\

\vspace{15mm}
Department of Physics,\\ The Ohio State University,\\ Columbus,
OH 43210, USA

\vspace {5mm}
mathur@mps.ohio-state.edu\\
turton.7@osu.edu
\vspace{4mm}

\end{center}

\vspace{1cm}

\begin{abstract}

If one attempts to add momentum-carrying waves to a black string then the solution develops a singularity at the horizon; this is a manifestation of the `no hair theorem' for black objects. However individual microstates of a black string do not have a horizon, and so the above theorem does not apply. We construct a perturbation that adds momentum to a family of microstates of the extremal D1-D5 string. This perturbation is analogous to the `singleton' mode
localized at the boundary of $AdS$; to leading order it is pure gauge in the $AdS$ interior of the geometry.

\end{abstract}

\thispagestyle{empty}

\newpage

\section{Introduction}

One of the great successes of string theory is its ability to reproduce the Bekenstein-Hawking entropy of black holes from a microscopic count of states~\cite{Susskind:1993ws,*Sen:1995in,Strominger:1996sh,*Callan:1996dv}. To resolve the information paradox~\cite{Hawking:1974sw,*Hawking:1976ra} however, we need to go beyond indirect counting and understand the {\it gravitational description} of individual black hole microstates. 

An early attempt to find such `black hole hair' was the construction of solutions describing travelling waves on a heterotic black string~\cite{Larsen:1995ss,Cvetic:1995bj,*Tseytlin:1996as,*Tseytlin:1996qg}, and a D1-D5 black string~\cite{Horowitz:1996th,*Horowitz:1996cj}.  
In each case the solution had a horizon. It turned out that adding the wave led to curvature singularities at the horizon 
in the form of infinite tidal forces
\cite{Kaloper:1996hr,*Horowitz:1997si} (see also~\cite{Banerjee:2009uk,Jatkar:2009yd}). One could regard this as an instance of the `no-hair theorem': there are no regular perturbations of a horizon. 

\begin{figure}[htbp]
\begin{center}
\includegraphics[width=10cm]{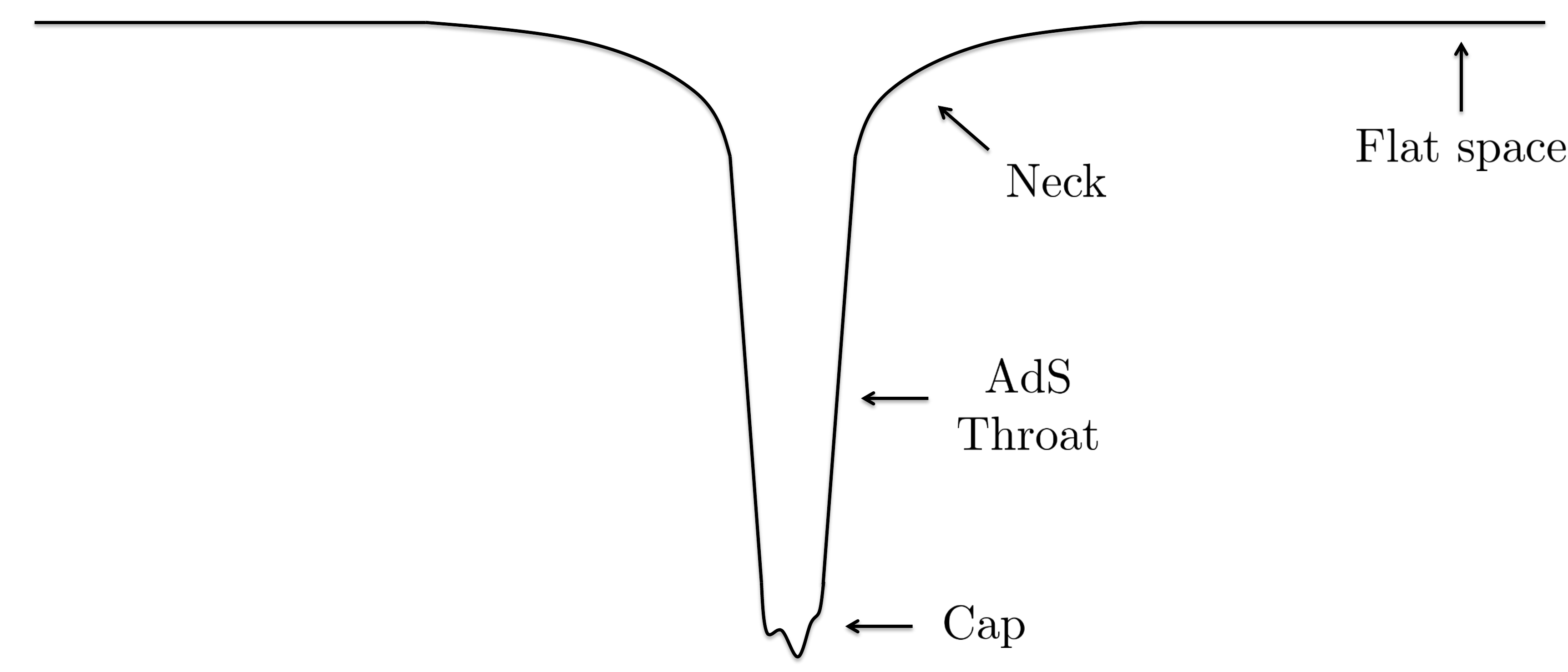}
\caption{Structure of two-charge D1-D5 geometries in the large $R_y$ limit.}
\label{fone}
\end{center}
\end{figure}

It turns out however that individual  states of the D1-D5 system do not have a traditional horizon. The simplest states are described by a semiclassical geometry with an AdS throat ending in a smooth cap~\cite{
Lunin:2001fv,Lunin:2001jy,Lunin:2002iz,*Kanitscheider:2007wq,*Skenderis:2008qn}.
This structure is depicted in Fig.\,\ref{fone} and described below. Since there is no horizon, we can ask the question: is there a perturbation of such D1-D5 geometries that is regular and carries momentum along the direction of the D1-D5 string?

In this paper we take a particular  class of D1-D5 microstate geometries, and give explicitly a perturbation that adds such momentum. The perturbation involves the directions along the compact torus $T^4$, and is thus the analog  of the `internal direction perturbations' that were studied in \cite{Larsen:1995ss,Cvetic:1995bj,*Tseytlin:1996as,*Tseytlin:1996qg,Horowitz:1996th,*Horowitz:1996cj}.  But since there is no horizon in the D1-D5 microstate, we evade the curvature singularity found in earlier works.

We work in type IIB string theory on $T^4\times S^1$. We wrap $n_1$ D1 branes on $S^1$ and $n_5$ D5 branes on $T^4\times S^1$, creating a D1-D5 bound state. 
If we take the $S^1$ radius $R_y$ to be the largest scale in the system, 
the background D1-D5 geometry has a long $AdS_3 \times S^3 \times T^4$ throat (Fig.\,\ref{fone}).

In this limit the D1-D5 bound state is described by a low energy 1+1 dimensional CFT.
This CFT has an `orbifold point' in moduli space~\cite{Seiberg:1999xz} that is the analog of a `zero coupling' point for the theory.
At the orbifold point the CFT is a sigma model  with target space $(T^4)^N/S_N$, where $N=n_1n_5$. 

The gravity perturbations we construct have a simple interpretation in the CFT: they describe excitations of the $U(1)$ currents in the CFT which arise from translation symmetry along the four directions of the $T^4$. (These $U(1)$ currents were studied in~\cite{Larsen:1999uk}.) 

At the orbifold point these currents have a simple expression in terms of the bosonic fields. 
The orbifold CFT is a symmetrized product of $N$ copies of a CFT with target space $T^4$. Let us label the different  copies by an index $r=1, \dots N$. Each copy has a bosonic oscillator for each of the four directions of the $T^4$; thus we have oscillator modes $\alpha_{-n}^{(r),i}$ with $i=1, \dots 4$. In terms of these modes, the $U(1)$ currents $J^i_{-n}$ are given by
\be
J^i_{-n} ~=~ \sum\limits_{r=1}^{N} \alpha_{-n}^{(r),i} \,. \nnm
\ee
Thus the perturbations we construct are the simplest excitations when considered from the viewpoint of the orbifold CFT: they are just the gravitational dual of a single bosonic oscillator mode (symmetrized among all the $N$ copies). 

For concreteness we first work with the simplest D1-D5 ground state -- the one with highest spin, which we denote by $|0\rangle_R$. 
The geometry sourced by $|0\rangle_R$ is known~\cite{
Balasubramanian:2000rt,Maldacena:2000dr}. 
We construct the dual perturbation for the CFT state
\be
|\Psi\rangle ~=~ J^i_{-n}|0\rangle_R \,. \nnm
\ee
Denoting the background metric and RR 2-form by $\bar{g}_{AB}, \, \bar{C}^{(2)}_{AB}$, the state $|\Psi\rangle$ is described by perturbed fields $\bar{g}_{AB}+ h_{AB}, \, \bar{C}^{(2)}_{AB}+ C_{AB}$. We obtain these perturbations $h_{AB}, C_{AB}$. 

In a recent paper~\cite{Mathur:2011gz} approximate forms of $h_{AB}, C_{AB}$ were found in two different regions of the geometry sourced by $|0\rangle_R$; matching in the overlap region was used to argue that a complete perturbation of the desired form should exist.  In this paper we find the full $h_{AB}, C_{AB}$ in closed form and extend the construction to a larger class of  background geometries. 

Let us discuss the  form expected of the perturbation. By the general ideas of Brown and Henneaux~\cite{Brown:1986nw}, the gravity duals of symmetry currents should be given by pure diffeomorphisms in an asymptotically $AdS$ geometry.\footnote{These ideas have been used by Strominger~\cite{Strominger:1997eq} and extended to a set of principles for general black holes by Carlip \cite{Carlip:1998wz,*Carlip:1999cy,*Carlip:1999db}.} Thus for each $U(1)$ currents, the gravity perturbation should be a diffeomorphism  (created  by  $T^4$ translations) in the  $AdS_3 \times S^3 \times T^4$ `throat' region of the geometry in Fig.\,\ref{fone}. But the perturbation cannot be a pure diffeomorphism {\it everywhere}, since the corresponding CFT excitation $J^i_{-n}$ adds an energy ${n\over R_y}$ to the state $|0\rangle_R$. We will observe that the perturbation we construct reduces to a diffeomorphism inside the `throat' region of the geometry, but is {\it not} a diffeomorphism in the `neck' (which is, in a sense, the boundary of the $AdS$ region).

Having the perturbation in closed form also permits us to examine it in a very different limit -- one in which $R_y$ is the smallest scale in the system, rather than the largest. In this limit the background D1-D5 geometry has the shape of a {\it ring}.
Interestingly, we find that the perturbation has structure over the entire region spanned by the ring (instead of being confined to the thickness of the ring). 

Finally, we extend the construction of the perturbation to a larger family of D1-D5 ground states -- the family of all D1-D5 ground states having $U(1)\times U(1)$ symmetry~\cite{Balasubramanian:2000rt,Maldacena:2000dr}. The energy of the gravity  perturbation in each case agrees with the excitation energy  expected for the corresponding  CFT state. 

This paper is organized as follows. In Section~\ref{sec:backgd} we introduce the background geometry 
created by $|0\rangle_R$ and the operator $J^i_{-n}$ in the D1-D5 CFT.
In Section~\ref{sec:pertn} we give the perturbation and analyze its properties. In Section~\ref{sec:gamma} we generalize the perturbation to the larger class of background geometries. In Section~\ref{sec:discussion} we discuss our results and future directions.

\section{\Large The background and the perturbation} \label{sec:backgd}

\subsection{The background geometry}

We work with type IIB string theory with the compactification
\be
M_{9, 1}~\r~ M_{4, 1}\times S^1\times T^4 \,.
\ee
We wrap $n_1$ D1 branes on $S^1$, $n_5$ D5 branes on $S^1\times T^4$, and consider the bound states of these branes. The fermions in the theory are periodic around the $S^1$, so the $1+1$ dimensional CFT in the $t, y$ directions is in the Ramond sector around the $y$ circle. 

This D1-D5 system is U-dual to an NS1-P system, where the D5 branes become a multiwound fundamental string along the $S^1$ and the D1 branes become momentum P carried as travelling waves on the string. The gravitational solution of such a string carrying waves is well known~\cite{Garfinkle:1990jq,*Callan:1995hn,*Dabholkar:1995nc,Lunin:2001fv}, and may be dualized back to obtain gravitational solutions describing the D1-D5 bound states~\cite{Lunin:2001fv,Lunin:2001jy}. We use the light-cone coordinates
\be
v = t-y \,, \qquad u = t+y 
\ee
constructed from the time $t$ and $S^1$ coordinate $y$. The different vibration profiles $\vec F(v)$ of the NS1P solutions leads to each D1-D5 solution being characterized by such a profile $\vec F(v)$. 

The generic state in this system is quantum in nature and the gravitational field it sources is not well-described by a classical geometry. This is simply because the momentum on the NS1 is spread over many different harmonics, with an occupation number of order unity per harmonic. One can however gain insight into the physics of generic solutions by starting with special ones, where all the momentum is placed in a few harmonics. In this case one can take coherent states for the vibrating NS1 string, which are then described by a classical profile function $\vec F(v)$ and which source classical geometries~\cite{Mathur:2005zp}. 

We start with the simplest geometry, which is obtained for the choice of profile function of the form
\be \label{eq:helix_profile}
F_1(v)=a\cos{v\over n_5 R_y}, \quad F_2(v)=a\sin{v\over n_{5} R_y}, \quad
F_3(v)=0, \quad F_4(v)=0 \,,
\ee
where $R_y$ is the radius of the $S^1$.
The geometry for this choice of $\vec F$ had arisen earlier in \cite{Cvetic:1996xz,Balasubramanian:2000rt,Maldacena:2000dr} and takes the form:
\bea
ds^2&=&-{1\over h}\left(dt^2-dy^2\right)+hf \left(d\theta^2+{dr^2\over r^2+a^2} \right) \cr
&& {}+h \left[ \left( r^2+{a^2Q^2\cos^2\theta\over h^2f^2} \right) \cos^2\theta d\psi^2
+\left(r^2+a^2-{a^2Q^2\sin^2\theta\over h^2f^2} \right)\sin^2\theta d\phi^2 \right] \cr
&& {} -{2aQ\over hf} \left(\cos^2 \theta dy d\psi+\sin^2\theta dt d\phi\right)
+
dz^i dz^i \,, 
\label{el} \\
C^{(2)}_{ty}&=&-{Q\over Q+f}  \,,
\qquad \qquad ~
C^{(2)}_{t\psi} ~=~ -{Qa\cos^2\theta\over Q+f}  \,,  \nonumber\\
C^{(2)}_{y\phi}&=&-{Qa\sin^2\theta\over Q+f} \,, 
\qquad\quad 
C^{(2)}_{\phi\psi} ~=~ Q\cos^2\theta+{Qa^2\sin^2\theta\cos^2\theta\over Q+f}   \label{selfdualfield}
\eea
where
\be
a={Q\over R_y}, \qquad f=r^2+a^2\cos^2\theta, \qquad h=1+{Q\over f} \,.
\ee

\subsection{The long AdS throat limit} \label{sec:long_thr}

As mentioned above, we will be interested in taking a limit of parameters where the geometry has the form depicted in Fig.\,\ref{fone}, where we have a long `throat' region where the geometry is locally $AdS$. To take this limit, we take $R_y$ to be large compared to $\sqrt{Q}$:
\be
\epsilon ~=~ 
{\sqrt{Q}\over R_y} ~\ll~ 1 \,.
\label{limitq}
\ee
We take $\e$ small enough that we get several units of the $AdS$ radius between the cap and the neck.

The different parts of the geometry (\ref{el}) then have the following structure (Fig.\,\ref{fone}) which is generic for two-charge geometries:

\begin{enumerate}[(i)]
	\item For $r\gg \sqrt{Q}$ we have approximately flat space.
	\item At $r\sim \sqrt{Q}$ we have the intermediate `neck' region.
	\item For $a \ll r \ll \sqrt{Q}$ we have the `throat', which is locally $AdS_3 \times S^3 \times T^4$.
	\item At $r\sim a$ we have the `cap'.
\end{enumerate}

In the case of the background \eq{el}, the cap is global $AdS_3$. 
We refer to the combined throat+cap as the `inner region'. If one decouples the inner region completely from the rest of the geometry, one obtains an asymptotically $AdS_3 \times S^3 \times T^4$ geometry which is dual to a 1+1 dimensional $\cN=(4,4)$ CFT.

\subsection{$U(1)$ currents in the D1-D5 orbifold CFT}

The D1-D5 bound state is described in the IR limit by a 1+1 dimensional sigma model, with base space $(t,y)$ and target space a deformation of the orbifold $(T^4)^N/S_N$, the symmetric product of $N$ copies of $T^4$.
The CFT has $(4,4)$ supersymmetry, and a moduli space which preserves this supersymmetry. It is conjectured that in this
moduli space there is an `orbifold point' where the target space is just the orbifold $(T^4)^N/S_N$ \cite{Seiberg:1999xz}.

The CFT with target space just one copy of $T^4$ is described by four real bosons $X^i$, four left-moving fermions $\psi^i$ and four right-moving fermions $\bar\psi^i$. The central charge is $c=6$. The complete theory with target space $(T^4)^N/S_N$ has $N$ copies of this $c=6$ CFT, with states that are symmetrized between the $N$ copies.  The orbifolding also generates `twist' sectors, which will be relevant later in this paper.

We use $r$ to label different copies of the $c=6$ CFT. At the orbifold point we note that each copy has four holomorphic $U(1)$ currents
\be
J^{(r),i} ~=~ \p X^{(r),i}
\ee
arising from translations in the four torus directions. The total CFT then has the $U(1)$ currents
\be
J^i ~=~ \sum\limits_{r=1}^{N} \p X^{(r),i} \,.
\ee
The modes of the current $J^{(r),i}$ are then simply the bosonic oscillator modes $\alpha_{-n}^{(r),i}$, and the modes of $J^i$ are
\be
J^i_{-n} ~=~ \sum\limits_{r=1}^{N} \alpha_{-n}^{(r),i} \,.
\ee
Based on the work of Brown and Henneaux~\cite{Brown:1986nw}, the gravity dual of 
\be
J^i_{-n} |0\rangle_R
\ee
should be a perturbation around the geometry \eq{el}, with the property that in the $AdS_3 \times S^3 \times T^4$ throat it reduces to a diffeomorphism. At the boundary of the $AdS_3 \times S^3 \times T^4$ region, the diffeomorphism should simply be that of a translation along the $T^4$ direction $i$, as a function of $v$~\cite{Mathur:2011gz}. In the next section we describe a perturbation with exactly these properties.

\subsection{The field equations for the perturbation}

We consider a perturbation given by the fields
\bea
g = \bar{g} + \hat{\e} \, h \,, \qquad C^{(2)} = \bar{C}^{(2)} + \hat{\e} \, C \,,
\eea
where $\hat{\e} \ll 1 $ is a small parameter.
From now on we single out a particular direction in the torus to work with,
\bea
z^1 ~\equiv~ z \,,
\eea
and take the perturbation to be given by 
\be \label{eq:ansatz}
h_{Az} \,, \quad C_{Az} \,.
\ee
To linear order in the perturbation, we  have $(F^{(3)})^2=0$, so that  the Ricci scalar $R$ vanishes. 
The equations for the perturbation are obtained from linearizing about the background the 10D equations ($M,N\ldots=1, \dots ,10$)
\bea
R_{MN}&=&{1\over 4}F^{(3)}_{MPQ}F^{(3)}_N{}^{PQ} \,, \label{qnine}\\
F^{(3)}_{MNP}{}^{;P}&=&0, \qquad F^{(3)}_{MNP}=\p_M C^{(2)}_{NP}+\p_N C^{(2)}_{PM}+\p_P C^{(2)}_{MN} \,.
\label{feqq}
\eea
We write 
\be
h_{Az}=A_A, ~~~C_{Az}=B_A, ~~~F_{AB}=\p_A A_B-\p_B A_A, ~~~G_{AB}=\p_A B_B-\p_B B_A \,.
\label{qten}
\ee
Then we find $R_{zA}~=~\h F_{AB}{}^{;B}$, and the Einstein equation (\ref{qnine}) reduces to
\be
F_{AB}{}^{;B} ~=~  \h G_{BC}F^{(3)}_A{}^{BC}  \,.
\label{qseven}
\ee
The field equation (\ref{feqq})  gives
\be
G_{AB}{}^{;B} ~=~ \h F_{BC}F^{(3)}_A{}^{BC}  \,
\label{qeight}
\ee
where the terms on the RHS arise from the connection term in the covariant derivatives. The equations (\ref{qseven}), (\ref{qeight}) can be separated by writing
\be
K_{AB}={1\over\sqrt{2}}\Big (F_{AB}+G_{AB}\Big ), ~~~~L_{AB}={1\over \sqrt{2}}\Big (F_{AB}-G_{AB}\Big ).
\ee
Following \cite{Mathur:2011gz} we work in an ansatz where we set 
\be \label{eq:torus_ansatz}
h_{Az} + C_{Az} ~=~ 0
\ee
which gives $K_{AB}=0$ and leaves us with the equation
\be \label{eq:eom}
L_{AB}{}^{;B} ~=~ - \h L_{BC}F^{(3)}_A{}^{BC}  \,.
\ee

\section{The perturbation and its properties} \label{sec:pertn}

\subsection{The perturbation}

We now present a perturbation which solves the linearized equations of motion \eq{eq:eom} on the background \eq{el}. The perturbation is given by the following fields: 
\bea
h_{vz} &=& e^{-i n \frac{v}{R_y}} \left( \frac{r^2}{r^2+a^2} \right)^{\!\!\frac{n}{2}} \frac{Q}{Q+f} \,, \cr
h_{rz} &=& e^{-i n \frac{v}{R_y}} \left( \frac{r^2}{r^2+a^2} \right)^{\!\!\frac{n}{2}} \frac{i \, a \, Q}{r(r^2 + a^2)} \,, \cr
h_{\psi z} &=&  e^{-i n \frac{v}{R_y}} \left( \frac{r^2}{r^2+a^2} \right)^{\!\!\frac{n}{2}} \frac{Q}{Q+f} 
\left( - a \cos^2 \th \right) , 
\label{eq:pertn}\\
h_{\phi z} &=&  e^{-i n \frac{v}{R_y}} \left( \frac{r^2}{r^2+a^2} \right)^{\!\!\frac{n}{2}} \frac{Q}{Q+f} 
\left( - a \sin^2 \th \right)  \, \nonumber
\eea
and
\be 
C_{Az} ~=~ - h_{Az} \,.
\ee
In the following section we will comment on the properties of this perturbation.  In the perturbation, $n$ takes the values
\be
n~=~1,2, 3, \dots
\ee
After quantization, negative values of $n$ will correspond to annihilation operators for the same modes that are described by positive $n$ (see \cite{Mathur:2011gz} for a more detailed discussion). The case $n=0$ describes a deformation that carries $U(1)$ charge; these charges were considered in \cite{Larsen:1999uk} but we will not focus on such solutions here.

We now comment on how this perturbation was derived. In~\cite{Mathur:2011gz}, an approximate construction of this perturbation was made following a method developed in~\cite{Mathur:2003hj}.
The method involves taking two limits of the background geometry, the `outer region' given by $r \gg a$ and the `inner region' given by $r \ll \sqrt{Q}$. 

The equations of motion may be solved in the two limits and the solutions matched in the region of overlap $a \ll r \ll \sqrt{Q}$.
In~\cite{Mathur:2011gz}, this procedure was followed to first order in this approximation scheme. By following the procedure to higher orders, and demanding a regular solution at each order, one is led to the above closed-form solution.

\subsection{Properties of the perturbation}

In Section \ref{sec:long_thr} we noted that when we take $R_y\gg \sqrt{Q}$ then we get a `long AdS throat'.  In this regime of parameters, we will see that the perturbation (\ref{eq:pertn}) has the following properties.  The perturbation: 
\begin{enumerate}[(i)]
	\item is normalizable at infinity;
	\item reduces to being a pure diffeomorphism in the $AdS$ throat+cap to leading order;
	\item is {\it not} pure gauge everywhere;
	\item is smooth in the `cap' region;
	\item has the correct spacetime dependence to carry the energy and momentum \\ required by the excitation $J^i_{-n}$.
\end{enumerate}

We shall demonstrate (i), (ii)  and (iv) explicitly; the other properties are easily verified.

\subsubsection{Normalizablility at infinity}

We first show that the perturbation is normalizable at infinity by examining the stress-energy tensor for the reduced gauge field ${1\over \sqrt{2}}(A-B)$, where $A, B$ were defined in (\ref{qten}).  To do this we examine the fall-off of $T_{00}$, the $tt$ component of the stress-energy tensor.

The background $F^{(3)}$ falls off as $1/r^3$ as $r \to \infty$ in an orthonormal frame. Its coupling to the field ${1\over \sqrt{2}}(A-B)$ then contributes at subleading order in $1/r$ and so to leading order at infinity the stress-energy tensor is that of a normal gauge field, 
\be
T_{AB} ~=~ L_{AC} L_{B}^{~\,C} - \frac{1}{4} g_{AB} \, L_{CD} L^{CD} \,.
\ee
We then find that
\be
T_{00} ~\sim~ {1\over r^{6}} \quad \mathrm{as} \quad r \to \infty \,.
\ee
Integrating $T_{00}$ over the remaining space directions then gives
\bea
\int T_{00} \, r^3 dr \, dy \, d\Omega_{3}   ~\sim~ {1\over r^{2}}
\eea
so we see that the perturbation is normalizable at infinity.

\subsubsection{Throat+cap region of background}

In order to show properties (ii) and (iv), we take the throat+cap limit of the background. This is obtained by taking the limit
\be
r ~ \ll ~ \sqrt{Q}
\label{qeleven}
\ee
in the full geometry (\ref{el}). Recall that we have taken $\epsilon={\sqrt{Q}\over R_y}\ll 1$. In the limit (\ref{qeleven}) we have ${Q\over Q+f}\r 1$. 

The background metric \eq{el} is singular when $f=0$, which occurs when $r=0$ and $\theta={\pi\over 2}$.  To show that the fields are regular at $r=0$ we must first change to coordinates in which the background is smooth at $r=0$. We thus make the spectral flow~\cite{Schwimmer:1986mf,Balasubramanian:2000rt,Maldacena:2000dr} transformation to the NS coordinates
\be
\psi_{NS} ~=~ \psi-{a\over Q}y, \quad \phi_{NS} ~=~ \phi-{a\over Q}t \,.
\label{spectral}
\ee
We then make the change of coordinates
\be \label{eq:primed_coords}
t'~=~{at\over Q} \,, \quad y' ~=~ {ay\over Q} \,, \quad r' ~=~ {r\over a} 
\ee
under which the 10D throat+cap metric becomes global $AdS_3 \times S^3 \times T^4$:
\bea
ds'^2&=& Q \left[ -(1+r'^2)dt'^2+{dr'^2\over 1+r'^2}+r'^2 dy'^2 + d\Omega_3^2 \right] +dz^i dz^i  \,
\label{aone}
\eea
where $d\Omega_3^2=d\theta^2+\cos^2\theta d\psi_{NS}^2+\sin^2\theta d\phi_{NS}^2$ is the metric on a round unit $S^3$.

\subsubsection{Pure gauge to leading order in the throat+cap}

We first show that to leading order in the $AdS$ throat+cap, the perturbation reduces to being a pure diffeomorphism accompanied 
by a gauge transformation
\be
C^{(2)}  \r C^{(2)} + \hat\epsilon \, d \Lambda \,.
\label{gaugeq}
\ee
When we go to a unit orthonormal frame, we find that the components $h_{\hat \psi \hat z}, h_{\hat \phi \hat z}$ are smaller than the other nonvanishing components by a factor ${a\over \sqrt{Q}}={\sqrt{Q}\over R_y}=\epsilon$. Thus we ignore these components, leaving
\bea
h_{vz} &=& e^{-i n \frac{v}{R_y}} \left( \frac{r^2}{r^2+a^2} \right)^{\!\!\frac{n}{2}} \,, \cr
h_{rz} &=& i e^{-i n \frac{v}{R_y}} \left( \frac{r^2}{r^2+a^2} \right)^{\!\!\frac{n}{2}} \frac{a Q}{r(r^2 + a^2)} \,.
\label{qone}
\eea
The gauge field also has two nontrivial components, since
\bea
C_{Az} &=&- h_{Az} \,.
\label{qtwo}
\eea
We now observe that the above metric and gauge field perturbations (\ref{qone}), (\ref{qtwo}) are generated by a diffeomorphism with parameter $\xi$ and a gauge transformation with parameter $\Lambda$, where 
\bea \label{eq:torus_diffeo}
\xi_z &=& i \frac{R_y}{n} e^{-i n \frac{v}{R_y}} \left( \frac{r^2}{r^2+a^2} \right)^{\!\!\frac{n}{2}} \,, \cr
\Lambda_z &=& -\xi_z \,.
\eea
The upper region of the throat, given by $r \gg a$, is analogous to the region near the boundary of $AdS$ after decoupling the throat from the flat asymptotics. We next observe that in this region, to leading order the above diffeomorphism becomes a translation in the $z$ direction as a function of $v$ only:
\bea \label{eq:torus_diffeo_asymp}
\xi_z &~\to~& i \frac{R_y}{n} e^{-i n \frac{v}{R_y}}  \,. 
\eea
As shown in~\cite{Mathur:2011gz}, this connects this perturbation, via the work of Brown and Henneaux, to the $U(1)$ current $J^i_{-n}$.

\subsubsection{Smoothness at the origin}

To show that the fields are smooth at the origin, we make the diffeomorphism with parameter $\xi$ and gauge transformation with parameter $\Lambda$ given by
\bea \label{eq:torus_diffeo_2}
\xi_z &=& -i \frac{R_y}{n} e^{-i n \frac{v}{R_y}} \left( \frac{r^2}{r^2+a^2} \right)^{\!\!\frac{n}{2}} \,, \cr
\Lambda_z &=& -\xi_z \,.
\eea
This removes the pure gauge part of the perturbation described above, including the component $h_{rz}$ in \eq{eq:pertn} which appeared potentially singular at $r=0$.
We next take the cap limit 
\be
r \ll \sqrt{Q}
\ee
and spectral flow to NS coordinates using \eq{spectral}.
This gives the cap fields
\bea
h_{vz} &=& e^{-i n \frac{v}{R_y}} \left( \frac{r^2}{r^2+a^2} \right)^{\frac{n}{2}} \left( - \frac{a^2}{2Q} - \frac{r^2}{Q} \right)  \,, \cr
h_{uz} &=& e^{-i n \frac{v}{R_y}} \left( \frac{r^2}{r^2+a^2} \right)^{\frac{n}{2}} \left( - \frac{a^2}{2Q} \right)  \,, \cr
h_{\psi z} &=& - a \, e^{-i n \frac{v}{R_y}} \left( \frac{r^2}{r^2+a^2} \right)^{\frac{n}{2}} \cos^2 \th \,,  \label{eq:innersolNS} \\
h_{\phi z} &=& - a \, e^{-i n \frac{v}{R_y}} \left( \frac{r^2}{r^2+a^2} \right)^{\frac{n}{2}} \sin^2 \th \, \nnm
\eea
and
\bea 
C_{Az} &=& - h_{Az} \,. \nnm
\eea
It can be checked that these fields are smooth at $r=0$.

\subsection{Perturbation in the black ring limit} \label{sec:ring}

\begin{figure}[h]
\begin{center}
\includegraphics[width=15cm]{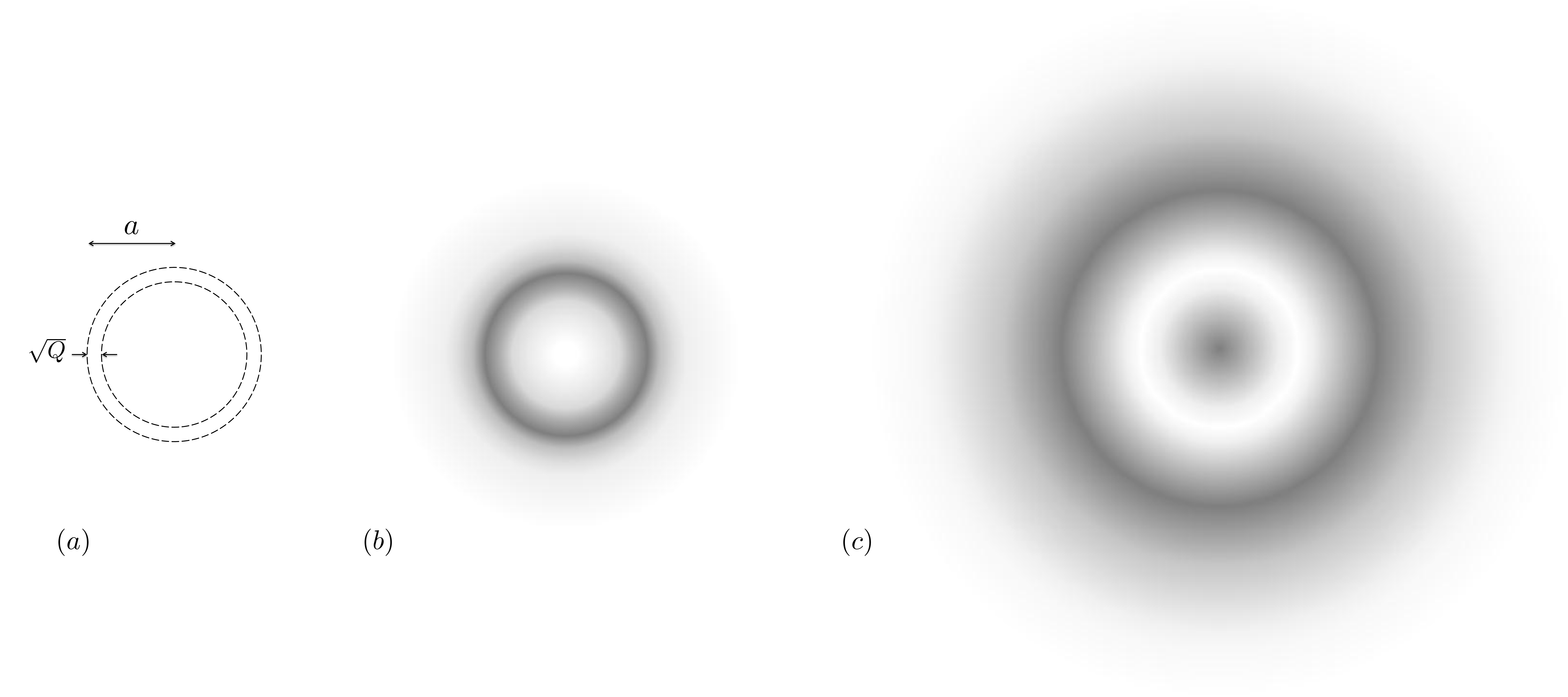}
\caption{(a) The black ring geometry obtained in the limit of small $R_y$ (b) The perturbation of \cite{Giusto:2006zi} was confined to the vicinity of the ring (c) The perturbation of the present paper is small at the ring, and spreads over a region that has the size of the entire ring.}
\label{fig_ring}
\end{center}
\end{figure}

In the above discussion we had taken the $S^{(1)}$ radius large compared to the curvature scale of the geometry: $R_y\gg \sqrt{Q}$. In this limit the solution (\ref{el}) takes the form  depicted in Fig.\,\ref{fone}. But we can also take the opposite limit: $R_y\ll \sqrt{Q}$. In this limit the geometry (\ref{el}) becomes a {\it ring}, depicted in Fig.\,\ref{fig_ring}\,(a). The ring has a `thickness' $\sqrt{Q}$, and a  radius $a={Q\over R_y}\gg \sqrt{Q}$. The metric is nontrivial within the thickness of the ring, and goes over to flat space when the distance from the ring axis becomes much larger than $\sqrt{Q}$. 

Since the perturbation (\ref{eq:pertn}) that we have constructed is given in closed form as a function of $R_y, Q$, we can examine it in the ring limit as well. A priori, we would anticipate one of two possibilities: (i) the perturbation may be confined to the ring thickness, dying away at distances much larger than $\sqrt{Q}$ from the ring axis (ii) the perturbation spreads over a much larger region of the size of the ring itself, controlled by the scale $a$. 

For comparison, in \cite{Giusto:2006zi} a momentum-carrying perturbation of the D1-D5 background (\ref{el}) was studied, obtaining a microstate of the D1-D5-P black ring~\cite{Elvang:2004rt,*Elvang:2004ds}. 
In that case the perturbation was of type (i). We depict this in Fig.\,\ref{fig_ring}\,(b). Interestingly, in the present case we find that our perturbation is of type (ii). To see this, note that the hierarchy of scales is now
\be
R_y ~~\ll~~ \sqrt{Q} ~~\ll~~ a \,.
\label{scalesbr}
\ee
Consider for example $h_{vz}$:
\bea
h_{vz} &=& e^{-i n \frac{v}{R_y}} \left( \frac{r^2}{r^2+a^2} \right)^{\frac{n}{2}} \frac{Q}{Q+f} \,.
\eea
At the axis of the ring, we have $r=0$, and we see that near this location  the perturbation vanishes as $r^n$. In fact the perturbation grows up to the scale $r \sim a$ before eventually falling off as $1/r^2$. So we see the perturbation does not have its main structure inside the `tube' $r \lesssim \sqrt{Q}$ but has structure on the scale of the ring itself. We depict this in Fig.\,\ref{fig_ring}\,(c).

The source of the difference between the two cases can be traced to the following fact. In the case studied in \cite{Giusto:2006zi} the perturbation had a wavenumber $k$ along the long axis of the ring, and by taking $k$ large one obtained a perturbation that fell off rapidly outside the thickness of the ring. Our present perturbation has no oscillations along the ring axis; thus increasing the level $n$ of $J^i_{-n}$ does not give us a high wavenumber along the ring.  So the perturbation for the $J^i_{-n}$ are analogous to the small $k$ case of \cite{Giusto:2006zi}. For small $k$ the perturbation of \cite{Giusto:2006zi} also spread over the scale $a$ of the entire ring.

\section{Extension to a larger class of backgrounds} \label{sec:gamma}

\subsection{The perturbation on conical defect metrics}

We have so far worked with the background sourced by $|0\rangle_R$, the Ramond ground state which is the spectral flow of the NS vacuum of the D1-D5 CFT. This geometry is one of a family of D1-D5 geometries studied in~\cite{Balasubramanian:2000rt,Maldacena:2000dr}. This family is characterized by a parameter
\bea
\g ~=~ 1/k \,, \qquad k ~=~ 1,2,\ldots,N  \,, \qquad N ~=~ n_1 n_5 \,.
\eea
The case $k=1$ corresponds to the state $|0\rangle_R$ which is described by the solution (\ref{el}). The other members of the family are 
obtained in the CFT  by acting on $|0\rangle_R$ with twist operators, for details see~\cite{Lunin:2001jy}. The resulting states have $N/k$ `component strings' of winding number $k$ each; each component string is generated by the action of a twist operator $\sigma_k$ that twists together $k$ copies of the $c=6$ CFT.

\begin{figure}[htbp]
\begin{center}
\includegraphics[width=11cm]{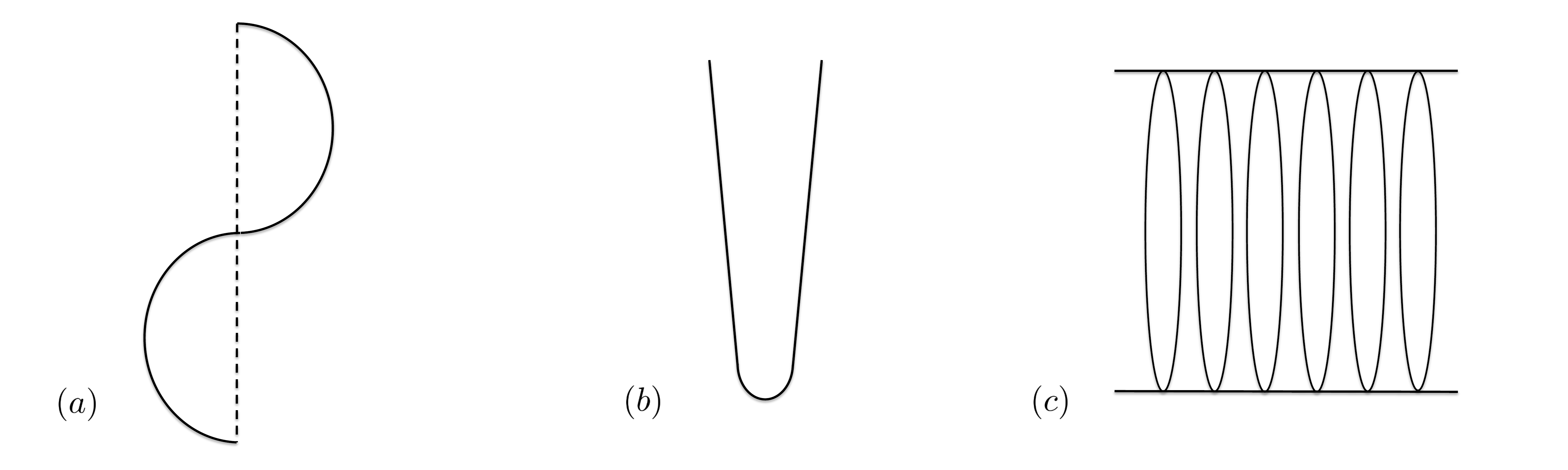}
\caption{The background geometry and corresponding CFT state for $k=1$. $(a)$ The NS1-P profile is in the lowest harmonic; $(b)$ the cap is global $AdS_3$; $(c)$ the dual CFT state consists of $N$ untwisted component strings.}
\label{fig:k=1}
\end{center}
\end{figure}

\begin{figure}[htbp]
\begin{center}
\includegraphics[width=11cm]{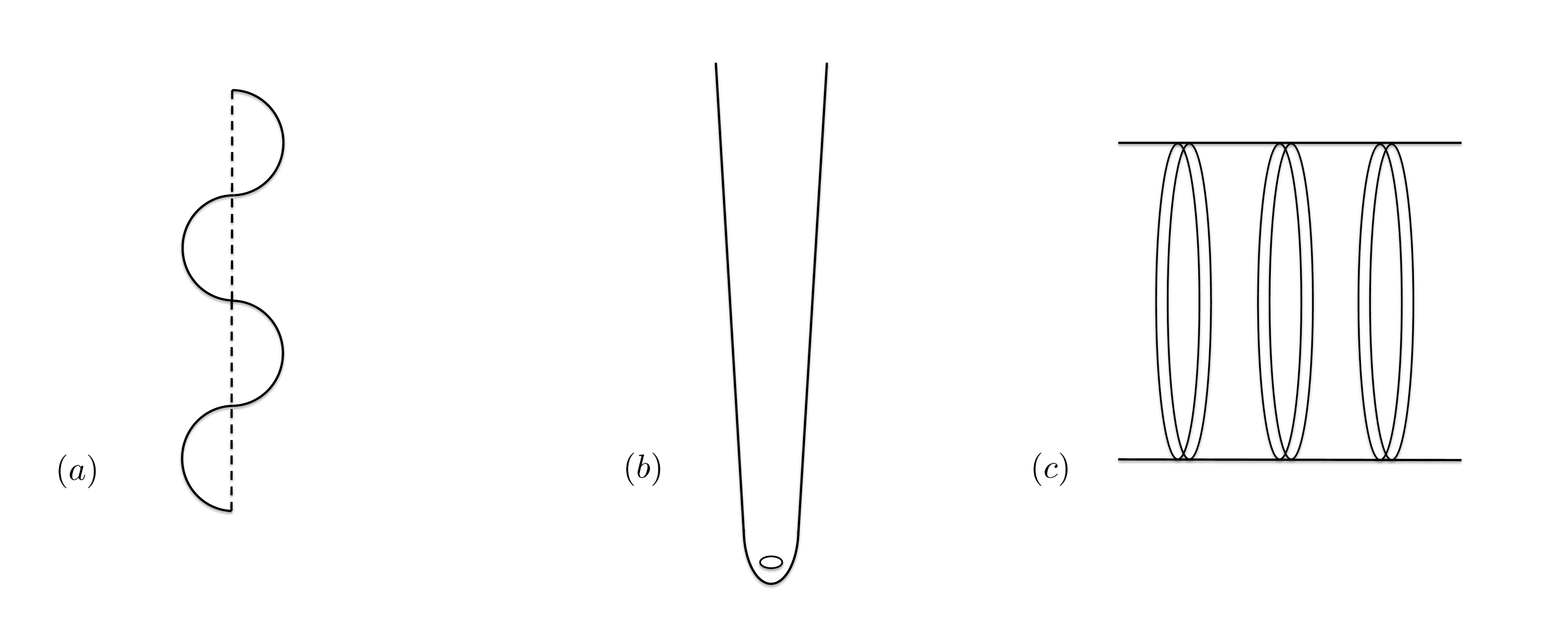}
\caption{The background geometry and corresponding CFT state for $k=2$. $(a)$ The NS1-P profile is in the 2nd harmonic; $(b)$ the throat is longer than for $k=1$, and there is a conical singularity along a curve in the cap; $(c)$ the dual CFT state consists of $N/2$ twisted component strings of length 2.}
\label{fig:k=2}
\end{center}
\end{figure}

We have seen that the gravitational solution for $k=1$ and has a smooth cap which is global $AdS_3$. 
For $k\neq 1$ however, the geometries have conical singularities. 
This is related to the fact that these solutions are U-dual to an NS1-P profile which runs over itself $k$ times. 
In Fig.\,\ref{fig:k=1} and Fig.\,\ref{fig:k=2} we qualitatively sketch the cases $k=1$ and $k=2$. In each figure, we first sketch the U-dual NS1-P profile, then the throat+cap region of the corresponding D1-D5 geometry, and finally the dual orbifold CFT state.


The gravity solutions for this family are characterized by a parameter $\gamma={1\over k}$, and are  given by~\cite{Balasubramanian:2000rt,Maldacena:2000dr}
\bea
ds^2 & = & -\frac{1}{h} (dt^2-dy^2) + h f \left( \frac{dr^2}{r^2 +
a^2\gamma^2} + d\theta^2 \right)
\nonumber \\
         &+& h \Bigl( r^2 +
\frac{a^2\gamma^2\,Q^2\,\cos^2\theta}{h^2 f^2} \Bigr)
\cos^2\theta d\psi^2  \nonumber \\
&+& h\Bigl( r^2 + a^2\gamma^2 -
\frac{a^2\gamma^2\,Q^2 \,\sin^2\theta}{h^{2} f^{2} }
\Bigr) \sin^2\theta d\phi^2  \nonumber \\
&-& \frac{2a\gamma\,Q }{hf}
(\cos^2\theta \,dy\,d\psi + \sin^2\theta \,dt\,d\phi)
+
dz^i dz^i \,, \label{mm} \\
C^{(2)}_{ty}&=&-{Q\over Q+f}  \,,
\qquad \qquad ~
C^{(2)}_{t\psi} ~=~ -{Qa \gamma \cos^2\theta\over Q+f}  \,,  \nonumber\\
C^{(2)}_{y\phi}&=&-{Qa \gamma \sin^2\theta\over Q+f} \,, 
\qquad\quad 
C^{(2)}_{\phi\psi} ~=~ Q\cos^2\theta+{Qa^2 \gamma^2 \sin^2\theta\cos^2\theta\over Q+f}   \label{selfdualfield_2}
\eea
where
\be \label{eq:params_gamma}
a={Q\over R_y}, \qquad f=r^2+a^2 \gamma^2 \cos^2\theta, \qquad h=1+{Q\over f} \,.
\ee

The perturbation \eq{eq:pertn} has a simple generalization to the case of $\g \neq 1$. We first write down the answer, and in the next subsection we explain how it arises as a natural generalization of the solution for the case of $\gamma=1$. The perturbation for general $\gamma={1\over k}$ is
\bea
h_{vz} &=& e^{-i n \frac{v}{R_y}} \left( \frac{r^2}{r^2+a^2 \gamma^2} \right)^{\!\!\frac{nk}{2}} \frac{Q}{Q+f} \,, \cr
h_{rz} &=& e^{-i n \frac{v}{R_y}} \left( \frac{r^2}{r^2+a^2 \gamma^2} \right)^{\!\!\frac{nk}{2}}
 \frac{ i \, a \gamma \, Q}{r(r^2 + a^2\gamma^2)} \,, \cr
h_{\psi z} &=&  e^{-i n \frac{v}{R_y}} \left( \frac{r^2}{r^2+a^2 \gamma^2} \right)^{\!\!\frac{nk}{2}} 
\frac{Q}{Q+f} \left[ -a\gamma \cos^2 \th \right] , \label{eq:MM_g_pertn} \\
h_{\phi z} &=&  e^{-i n \frac{v}{R_y}} \left( \frac{r^2}{r^2+a^2 \gamma^2} \right)^{\!\!\frac{nk}{2}} 
\frac{Q}{Q+f} \left[ -a\gamma \sin^2 \th \right] \, \nnm
\eea
and
\bea 
C_{Az} &=& - h_{Az} \,, \nnm
\eea
where here $f$ is given in \eq{eq:params_gamma}.

\subsection{Derivation of the generalization}

When $\gamma \neq 1$ the `throat+cap' of the solution \eq{mm} is an orbifold of $AdS_3\times S^3$ (we ignore the $T^4$ for this discussion). This can be seen as follows. The throat+cap region is obtained by setting $r\ll \sqrt{Q}$, which gives $h=1+{Q\over f}\approx{Q\over f}$. In this limit the metric (\ref{mm}) is
\bea
ds^2&=& -{f\over Q}(dt^2-dy^2)~+~Q\Big ({dr^2\over r^2+a^2\gamma^2}+d\theta^2\Big )\nn
&& ~+~Q \cos^2\theta d\psi^2~+~Q\sin^2\theta d\phi^2\nn
&&~-~2 a \gamma \cos^2\theta dyd\psi ~-~2 a \gamma \sin^2\theta dt d\phi \,.
\eea
The metric can be diagonalized by introducing the coordinates
\be
\t t = {a \gamma\over Q} t, ~~~\t y={a\gamma\over Q} y, ~~~\t r={r\over a\gamma}, ~~~\t\theta=\theta, ~~~\t\psi=\psi-\t y, ~~~\t\phi=\phi-\t t
\label{qfour}
\ee
in terms of which the metric  becomes
\be
ds^2~=~Q \left[ -(\t r^2+1) d\t t ^2+\t r^2 d\t y^2+{d\t r^2\over \t r^2+1} ~+~ d\t\theta^2+\cos^2\theta d\t\psi^2+\sin^2\theta d\t\phi^2 \right]
\label{qthree}
\ee
which is (locally) the metric of $AdS_3\times S^3$. However, note that the identifications on the initial variables $y, \psi, \phi$ were
\be
y\r y+2\pi R_y \,,  \qquad \psi\r \psi+2\pi \,, \qquad \phi\r \phi+2\pi \,.
\ee
In the new variables, these become
\be
\left\{ \t y\r \t y +2\pi \gamma, ~~ \t\psi\r\t\psi-2\pi \gamma \right\} , \quad~~ \t\psi\r\t\psi+2\pi \,, \quad~~ \t\phi\r \t\phi+2\pi
\ee
where the first identification has a joint displacement of both $\t y$ and $\t\psi$. This identification has a fixed point where both the $\t y$ and $\t\psi$ circles collapse to zero size, which happens at $r=0, ~\t\theta={\pi\over 2}$. This is just the curve $f=0$ in the geometry (\ref{mm}). Thus this curve (which has the shape of a circle) is the location of an orbifold singularity. The singularity is a conical defect created by identifying an angular variable by $2\pi \gamma=2\pi/k$ instead of by $2\pi$. 
 
Since the metric (\ref{qthree}) has the same $AdS_3\times S^3$ form as in the case $\gamma=1$, we can locally solve the equations of motion by the perturbation (\ref{eq:pertn}) that we had for the case $\gamma=1$. The change of coordinates implies that the functions in the perturbation will come with factors of the form
\bea
e^{-i \widetilde{n} \widetilde{v} } \left( \frac{\widetilde{r}^2}{\widetilde{r}^2+1} \right)^{\!\!\frac{\widetilde{n}}{2}} \,.
\eea
We now rewrite this expression in the original coordinates of the metric (\ref{mm}), getting
\bea \label{eq:orig}
e^{-i \widetilde{n} \gamma \frac{v}{R_y} } \left( \frac{r^2}{r^2+ a^2\gamma^2} \right)^{\!\!\frac{\widetilde{n}}{2}} \,.
\eea
At infinity the momentum along $y$ must be quantized in units of $1/R_y$, so we must impose
\bea
\widetilde{n} \gamma &=& n \,, \qquad n = 1,2,3,\ldots
\eea
Then \eq{eq:orig} becomes
\bea
e^{-i n \frac{v}{R_y} } \left( \frac{r^2}{r^2+ a^2\gamma^2} \right)^{\!\!\frac{n k}{2}} \,
\eea
where we have written the power on the $r$ dependence in terms of $k=1/\gamma$ as before. Substituting this expression and replacing $a \to a \gamma$ where appropriate in \eq{eq:pertn} gives the generalized perturbation \eq{eq:MM_g_pertn}. We have verified that \eq{eq:MM_g_pertn} solves the perturbation equations around the metric \eq{mm}.

\subsection{The CFT excitation}

Let us look at the excitation in the orbifold CFT corresponding to the perturbation (\ref{eq:MM_g_pertn}). We imagine quantizing the mode (\ref{eq:MM_g_pertn}), and taking a single quantum of the perturbation. The quantum has energy $E=\omega={n\over R_y}$, which corresponds to an excitation at level $1$ in the CFT.  Since we have a single supergravity quantum, we expect that only one component string will be excited (see \cite{Lunin:2001jy} for more details on relating gravity quanta to CFT excitations). 

Denoting the Ramond vacuum of the $r$th component string by $|0_r\rangle_R$, the perturbation in the CFT with these characteristics is given by
\be
|\Psi\rangle ~=~  \sum_{r=1}^{N\over k} \alpha_{-nk}^{(r), i} \, \bigg( |0_1\rangle_R \otimes \cdots \otimes |0_{N\over k}\rangle_R  \bigg) \,.
\ee
Here we have a single bosonic oscillator in mode $nk$ on one of the component strings (and we have symmetrized over all the component strings). Since the CFT with twisted cycles lives in a `box' of length $2\pi k R_y$, the oscillator at level $nk$ carries energy ${n\over R_y}$, as desired.

Note that there are many other states possible at this energy level, since we are at level $nk$ on the twisted component string. Construction of the other states at this level is a more complicated task, which we hope to address elsewhere.

\newpage

\section{Discussion} \label{sec:discussion}

The gravitational description of all two-charge D1-D5 states can be understood. This is possible because  the D1-D5 system can be mapped by dualities to NS1-P, which is a string carrying travelling waves. The metrics for such strings are known~\cite{Garfinkle:1990jq,*Callan:1995hn,*Dabholkar:1995nc,Lunin:2001fv}, and dualizing back we get the metrics for D1-D5 bound states~\cite{Lunin:2001fv,Lunin:2001jy}.
There is however no known construction which gives all three-charge D1-D5-P solutions in a similar way. Instead, families of D1-D5-P states have been constructed by diverse methods
\cite{
Giusto:2004id,*Jejjala:2005yu,
Bena:2004de,*Balasubramanian:2006gi,*Bena:2007kg,*Bena:2011uw,
Giusto:2011fy,*Giusto:2012gt,*Vasilakis:2012zg}. In some cases, the gravity solutions have been identified with their counterparts in the orbifold CFT.

In this paper we have taken a particular set of CFT states $J^i_{-n}|0\rangle_R$ and constructed their gravity duals. These states are special in the following sense.  Recall that the CFT on $N$ D3 branes is $U(N)$ gauge theory, which splits into an $SU(N)$ factor (dual to $AdS_5\times S^5$~\cite{Maldacena:1997re}) and a $U(1)$ factor which is dual to a singleton~\cite{Dirac:1963ta,*Flato:1978qz} mode localized at the `boundary of $AdS$'~\cite{Witten:1998qj}. Similarly, in the D1-D5 case we have modes localized at the boundary of $AdS_3$~\cite{Gunaydin:1986fe,deBoer:1998ip} which are dual to CFT excitations created by the chiral algebra of the CFT. In particular the $U(1)$ chiral currents $J^i_{-n}$ generate such boundary modes, which we have studied in the present paper.

The background we work with is not, however, the $AdS$ space but the entire asymptotically flat geometry sourced by the D1 and D5 branes. Here the boundary of $AdS$ is naturally bounded by a `neck' region that leads to the asymptotically flat part of spacetime. Following the general arguments of Brown and Henneaux, we expect the singleton mode to be a diffeomorphism in the $AdS$ region. We find that our perturbation is pure gauge to leading order in the $AdS$ region, and nontrivial in the neck; this allows it to carry the required energy of the CFT excitation. 
	
The black string waves of \cite{Larsen:1995ss,Cvetic:1995bj,*Tseytlin:1996as,*Tseytlin:1996qg,Horowitz:1996th,*Horowitz:1996cj} were found to be singular at the horizon; the same was found of the corresponding modes in~\cite{Banerjee:2009uk} when examined at quadratic order~\cite{Jatkar:2009yd}. In our case the perturbation is given by smooth functions on a smooth geometry with no horizon, so we do not expect any such singularity.

The state in the orbifold CFT corresponding to this perturbation is the simplest possible excitation in the CFT. The orbifold CFT has (on each of the $N=n_1n_5$ copies) four free bosons and four free fermions. The perturbation we have constructed corresponds to exciting one bosonic oscillator mode $\alpha^i_{-n}$ (symmetrized  over the $N$ copies of the CFT).  

It is interesting to compare the CFT excitation for $J^i_{-n}$ with the excitation that corresponds to the Virasoro generators $L_{-n}$. 
 In terms of the bosonic oscillators $\alpha_{-n}^{(r),i}$ of the $r$th copy of the CFT, $J^i_{-n}$ is given by
\be
J^i_{-n} ~=~ \sum\limits_{r=1}^{N} \alpha_{-n}^{(r),i} \,
\ee
while $L_{-n}$ is given by
\be
L_{-n} ~=~ \sum\limits_{r=1}^{N} \alpha_{-n-m}^{(r),i} \alpha_{m}^{(r),i} \,.
\ee
Note however that acting twice with $J^i_{-n}$ is very different from acting with $L_{-n}$, since
\be
J^i_{-n_1} J^i_{-n_2}~=~ \left( \sum\limits_{r=1}^{N} \alpha_{-n_1}^{(r),i}  \right) \left( \sum\limits_{s=1}^{N} \alpha_{-n_2}^{(s),i} \right).
\ee
so that we have a very different correlation between the `copy index' of the two oscillators. This difference is analogous to the difference in the D3 brane case between the operators $(\tr F^4)$ and $(\tr F^2)^2$. 

Our solutions are new three-charge perturbations, which depend on the light-cone coordinate $v$. They are not generated by simply adding a wave to the background geometry in the way we add a wave to a string.
The latter approach needs a null Killing vector along which we must translate \cite{Garfinkle:1990jq}, and there is no such Killing vector for the background D1-D5 microstate solution (\ref{el}). Instead, there is a detailed structure of the perturbation in the cap region which ensures its regularity. 

There are many possible directions in which to continue this research. One is to construct the perturbation in this paper to quadratic order, and potentially to a fully nonlinear solution. 
Another possible direction is to generalize the construction of this paper to other background geometries, for example known classes of three-charge extremal and non-extremal solutions~\cite{Giusto:2004id,*Jejjala:2005yu}. It would also be interesting to see how our solutions relate to the analysis of~\cite{Bena:2011dd}.

\section*{Acknowledgements}

We thank Steve Carlip, Atish Dabholkar, Stefano Giusto, Murat Gunaydin, Dileep Jatkar, Rodolfo Russo and Yogesh Srivastava for helpful discussions. 
This work was supported in part by DOE grant DE-FG02-91ER-40690.

\newpage


\providecommand{\href}[2]{#2}\begingroup\raggedright\endgroup

\end{document}